\documentclass{article}
\usepackage[margin=1.25in]{geometry}
\usepackage{comment}
\usepackage{subcaption}

\usepackage{csvsimple,longtable,booktabs}

\usepackage{csvsimple,longtable,booktabs}
\usepackage{graphicx}
\usepackage{xcolor}

\usepackage{amsmath,amsthm,amsfonts}

\usepackage{amsmath,amsthm,amsfonts}
\usepackage[round]{natbib}

\usepackage{bm}
\usepackage{hyperref}

\begin{document}

\newcommand{\R}{\mathbb{R}}
\newcommand{\Cov}{\mbox{Cov}}
\newcommand{\veps}{\varepsilon}

\begin{center}
{\LARGE Vecchia Approximations and Optimization for \\ \vspace{6pt} Multivariate Mat\'ern Models}
\vspace{12pt}

{\large Youssef Fahmy, Joseph Guinness}

\vspace{12pt}

{\large \textit{Cornell University, Department of Statistics and Data Science} }
\vspace{12pt}

\textbf{Abstract}
\end{center}
We describe our implementation of the multivariate Mat\'ern model for multivariate spatial datasets, using Vecchia's approximation and a Fisher scoring optimization algorithm. We consider various pararameterizations for the multivariate Mat\'ern that have been proposed in the literature for ensuring model validity, as well as an unconstrained model. A strength of our study is that the code is tested on many real-world multivariate spatial datasets. We use it to study the effect of ordering and conditioning in Vecchia's approximation and the restrictions imposed by the various parameterizations. We also consider a model in which co-located nuggets are correlated across components and find that forcing this cross-component nugget correlation to be zero can have a serious impact on the other model parameters, so we suggest allowing cross-component correlation in co-located nugget terms.

\section{Introduction}

We attempt to understand the complexities of the Earth system by measuring and modeling many variables that interact on a continuum of spatial-temporal scales. For example, modern climate models include dozens of variables evolving in concert over space and time. In this paper, we analyze elemental data from a soil monitoring network in France \citep{saby2009multivariate}, elemental data from a well-water monitoring program in Bangladesh \citep{kinniburgh2001arsenic}, multispectral data from the GOES-16 satellite (maintained by NASA and NOAA), and the difference between forecasted and actual pressures and temperatures in the Pacific Northwest \citep{eckel_mass_2005}. 

Gaussian processes have become a workhorse for statistical analysis of spatial and spatial-temporal data.  A multivariate Gaussian process with $p$-components is a random vector function $Z(x) =(Z_{1}(x),\dots, Z_p(x))^{\top} \in \mathbb{R}^p$ indexed by a set $D \subset \mathbb{R}^d, d \geq 1,$ such that for any $x_1,\dots, x_n \in D,$ the random vector $(Z(x_1)^{\top},\dots, Z(x_n)^{\top})^{\top} \in \mathbb{R}^{np}$ has a multivariate normal distribution. The process is completely specified by its mean function $\mathbb{E}[Z(x)]$ and covariance function $\{\Cov(Z_i(x), Z_j(x'))\}_{i,j = 1}^p$. When the covariances depend only on the separation lag $h$--that is, the process is covariance-stationary--we write $C_{ij}(h)$ for $\Cov(Z_i(x+h), Z_j(x))$, which is referred to as a cross covariance function. \cite{genton2015cross} provide a thorough review of cross covariance functions.

Following the popularity and success of the Mat\'ern model for univariate spatial data, a multivariate version has been proposed and studied \citep{gneiting2010matern,apanasovich2012valid,emery_porcu_white_2022}. The general form of the multivariate Mat\'ern model is given by
\begin{align*}
   C_{ij}(h)= \sigma_{ij}M(h|\nu_{ij},\alpha_{ij})\,\,\,\,\,\,{h}\in \mathbb{R}^d, \,\,\,i,j = 1,\dots, p
\end{align*}
where
\begin{align*}
    M(h|\nu,\alpha) = \frac{1}{2^{\nu-1}\Gamma(\nu)}( \lVert h\rVert/\alpha)^{\nu}K_{\nu}( \lVert h\rVert /\alpha)
\end{align*}
with $K_{\nu}$ being the modified Bessel function of the second kind of order $\nu$. 
The parameter $\sigma_{ij}$ is the covariance between co-located observations from components $i$ and $j$. We refer to the parameters using the following terminology: $\sigma_{ij} $ is a cross covariance parameter, $\alpha_{ij}$ is a cross range parameter, and $\nu_{ij}$ is a cross smoothness parameter. To be consistent with how the Mat\'ern is parameterized in our existing software, our $\alpha_{ij}$ parameters are ranges, whereas they are inverse ranges in \cite{gneiting2010matern}, \cite{apanasovich2012valid}, and \cite{emery_porcu_white_2022}. When $i=j$, we refer to them as marginal parameters. Note that this model implies the symmetry $ C_{ij}(h) =C_{ji}(h)$ which need not hold in general. \cite{li2011approach} and \cite{qadir2021flexible} proposed methods for modeling asymmetries. 
The univariate Mat\'ern  model $C_{ii}(h) = \sigma_{ii}M(h|\nu_{ii}, \alpha_{ii})$ provides a valid (i.e nonnegative definite) second-order structure for the marginal process $Z_i(x)$ as long as the marginal parameters $\sigma_{ii}, \alpha_{ii} ,\nu_{ii} $ are positive. Additional conditions are needed on the  cross parameters $\sigma_{ij}, \alpha_{ij}$ and $\nu_{ij}$ to ensure that the multivariate Mat\'ern  model is valid for the multivariate process $Z(x)$. We discuss these in the next section.

\cite{kleiber2017coherence} studied the properties of various multivariate spatial models, including separable models, kernel convolution models, the linear model of coregionalization, and the multivariate Mat\'ern. He found that the multivariate Mat\'ern is sufficiently flexible in that it allows the high frequency coherence to exhibit a range of behaviors, depending on the parameter settings. Loosely speaking, the coherence between two components is the correlation between the linear combination of each component and a sinusoidal function; it measures correlation between components at a particular frequency of variation. \cite{qadir2021semiparametric} demonstrated that further improvements in flexibility can be achieved with semiparametric models. 

Over the past decades, spatial statisticians have produced a mountain of literature on the topic of estimating covariance parameters, especially on the problem of computing estimates when the dataset is very large. This work has led to various software packages, including \verb!INLA! \citep{INLA,lindgren2011explicit}, \verb!fields! \citep{fields}, \verb!RandomFields! \citep{RandomFields}, \verb!spBayes! \citep{spBayes}, \verb!spNNGP! \citep{spNNGP}, \verb!LatticeKrig! \citep{LatticeKrig}, \verb!FRK! \citep{FRK}, \verb!exageostat! \citep{abdulah2018exageostat}, \verb!GpGp! \citep{guinness2018gpgp}, \verb!GPvecchia! \citep{GPvecchia}, and \verb!GeoModels! \citep{GM2022}, to name a few. Most of the research and software development is focused on the univariate case, with the exception of \verb!RandomFields!, \verb!exageostat!, and \verb!spBayes!, which are capable of fitting bivariate Mat\'ern models, and \verb!GeoModels!, which has a number of bivariate spatial models.

Clearly, there is a need for reliable software capable of fitting multivariate spatial models with two or more components to large datasets. In this work, we report on extending the \verb!GpGp! R package \citep{guinness2018gpgp} to handle the multivariate Mat\'ern model, demonstrate its capabilities on several datasets, and explore the implications of various proposed sufficient conditions on multivariate Mat\'ern parameters. \verb!GpGp! implements Vecchia's Gaussian process approximation \citep{vecchia1988estimation}, along with improvements to the approximation \citep{guinness2018permutation}, and likelihood optimization procedures that efficiently compute and make use of the gradient and Fisher information \citep{guinness2021gaussian}. The application of Vecchia's approximation is agnostic to the covariance function, which has allowed for the implementation of more than 25 covariance models in \verb!GpGp! at the time of writing (package version 0.4.0), including isotropic, geometrically anisotropic, nonstationary, and spatial-temporal models on Euclidean spaces and spheres. \verb!GpGp! has enjoyed success in spatial data competitions, including being used by the winning team in the first KAUST spatial data analysis competition \citep{huang2021competition}, and by the winners of the multivariate spatial data analysis section of the second competition \citep{abdulah2022second}.

Adding the multivariate Mat\'ern model to \verb!GpGp! presents difficulties that go far beyond the normal challenges of implementing a typical univariate covariance function. As opposed to the univariate Mat\'ern, whose parameters must simply be positive in order for the model to be valid, known sufficient conditions are more complicated, so some care must be taken when enforcing them. The parameter space is also large; our formulation of the model, which allows for correlated nuggets, has $2p(p+1)$ parameters. Depending on the dataset, many of the parameters--or combinations thereof--are not well identified. In short, it is a nasty optimization problem. In order to quickly maximize the likelihood, one has to take large steps through a high dimensional space fraught with \verb!Error!s, \verb!Inf!s, and \verb!NaN!s. R's \verb!optim! function does not cut it. In addition, as with all Vecchia approximations, decisions must be made about how to order the observations and select neighbors, which is complicated by the fact that the multivariate component of the data is usually categorical rather than numeric.

Our major contribution is the software we provide for fitting multivariate Mat\'ern models using Vecchia's approximation and a Fisher scoring algorithm. However, the software allows us to explore the behavior of the multivariate Mat\'ern model on various datasets. Our findings are generally consistent with those of other authors; more flexible conditions on the parameters tend to give better fits, but there are diminishing returns on added flexibility. 
Perhaps our most interesting modeling finding is that allowing the nugget term to be correlated across components can have a large impact on the estimates of the other covariance parameters. 

Section \ref{model_section} reviews the multivariate Mat\'ern model and its various parameterizations. Section \ref{vecchia} outlines Vecchia's approximation for multivariate spatial data. Section \ref{optimization} includes some notes on the optimization procedures. Section \ref{data} describes the datasets. Section \ref{results} presents the results. We conclude with a discussion.

\section{Multivariate Mat\'ern Parameterizations}\label{model_section}

We model the responses as
\begin{align*}
Y_i(x) = \mu_i + Z_i(x) + \veps_{i}(x),
\end{align*}
where $\mu_i$ is component-specific mean, $Z_i$ is a multivariate Mat\'ern process, and $\veps_i(x)$ is a nugget term with covariances
\begin{align*}
\Cov( \veps_i(x+h), \veps_j(x) ) = \tau_{ij} \bm{1}(h = 0).
\end{align*}
In other words, we assume constant mean within each component, and we add a nugget term but allow the nugget to be correlated across components. The $p\times p$ matrix formed by the $\tau_{ij}$ parameters must be positive definite. We parameterize the cross nugget variances as $\tau_{ij} = (\tau_{ii}\tau_{jj})^{1/2}S_{ij},$ where $S$ is a correlation matrix.  

\cite{gneiting2010matern} provided necessary and sufficient validity conditions for the bivariate Mat\'ern model parameters. These conditions define the \emph{full bivariate model}. For three or more components, necessary and sufficient conditions on the parameters are not known, though various authors have proposed sufficient conditions, some of which we explore here.

\subsection{Parsimonious Model}

\cite{gneiting2010matern} proved that the multivariate Mat\'ern model is valid for $p \geq 2$ if the following conditions hold for every $i$ and $j$:
\begin{enumerate}
\item $\alpha_{ij} = \alpha$ (common marginal and cross ranges)
\item $\nu_{ij} = (\nu_{ii} + \nu_{jj})/2$
\item 
$ {\sigma_{ij}}= (\sigma_{ii}\sigma_{jj})^{1/2} V_{ij}\frac{\Gamma(\nu_{ii} + d/2)^{1/2}}{\Gamma(\nu_{ii})^{1/2}}\frac{\Gamma(\nu_{jj} + d/2)^{1/2}}{\Gamma(\nu_{jj})^{1/2}} \frac{\Gamma\{(\nu_{ii}+ \nu_{jj})/2\}}{\Gamma\{(\nu_{ii}+ \nu_{jj})/2 + d/2\}}$ where $V$ is a correlation matrix. 

\end{enumerate}
These conditions define their \emph{parsimonious model}. If we define $\rho_{ij}:= {\sigma_{ij}}/{(\sigma_{ii}\sigma_{jj})^{1/2}},$ then condition $3$ implies
\begin{align*}
  |\rho_{ij}|\leq \frac{\Gamma(\nu_{ii} + d/2)^{1/2}}{\Gamma(\nu_{ii})^{1/2}}\frac{\Gamma(\nu_{jj} + d/2)^{1/2}}{\Gamma(\nu_{jj})^{1/2}} \frac{\Gamma\{(\nu_{ii}+ \nu_{jj})/2\}}{\Gamma\{(\nu_{ii}+ \nu_{jj})/2 + d/2\}}
\end{align*}
which reduces to  $|\rho_{ij} | \leq \frac{(\nu_{ii}\nu_{jj})^{1/2}}{(\nu_{ii} + \nu_{jj})/2}$
 when $d = 2$.
In the bivariate case, condition $3$ provides a complete characterization of $\rho_{ij}$ when $1$ and $2$ hold. In our software, all correlation matrices, such as $V$ here, use a Cholesky-based parameterization \citep{Pinheiro96unconstrainedparameterizations}. All parameters that must be positive use an exponential/log link.

\subsection{Flexible-A Model}

\cite{apanasovich2012valid} provide a different set of sufficient conditions in the $p \geq 2$ case which do not require a common range parameter or the restriction that $ \nu_{ij} = ({\nu_{ii} + \nu_{jj}})/{2}$: 
\begin{enumerate}
    \item
 $ \nu_{ij} = \frac{\nu_{ii} + \nu_{jj}}{2} + \Delta_A(1 - A_{ij})$ where $\Delta_A\geq 0$  and $A$ is a positive correlation matrix,
     \item
    $(\alpha_{ij}^{-2})_{i,j = 1}^p$ is conditionally negative semidefinite,
       \item
       $
 \sigma_{ij} = (\sigma_{ii}\sigma_{jj})^{1/2}V_{ij}(u_{ii}u_{jj})^{-1/2}u_{ij}$,
where
\begin{align*}
         u_{ij} = \alpha_{ij}^{2\Delta_A + \nu_{ii} + \nu_{jj}}\Gamma(\nu_{ij})\Gamma\{(\nu_{ii} + \nu_{jj})/2 + d/2\}/\Gamma(\nu_{ij} + d/2),
\end{align*}
and $V$ is a correlation matrix.  
\end{enumerate}
We will refer to the model defined by these conditions as the \emph{Flexible}-$A$ model. As suggested by \cite{apanasovich2012valid}, we ensure condition $2$ holds by parameterizing
\begin{align*}
     \alpha^{-2}_{ij}  = \frac{\alpha^{-2}_{ii} + \alpha^{-2}_{jj}}{2}  + \Delta_B(1 - B_{ij})
\end{align*}
where $\Delta_B\geq 0$  and $B$ is a positive correlation matrix. In the bivariate case, $A$ and $B$ become redundant.

The conditions of the Flexible-A model are not necessary, so in the bivariate case the full model of \cite{gneiting2010matern} is less restrictive. However, the conditions of the full model are more complicated to enforce,  so it 
is still worthwhile to evaluate the performance of the Flexible-A model on bivariate datasets. \cite{apanasovich2012valid} illustrate on the same bivariate weather dataset considered by  \cite{gneiting2010matern} that both models obtain similar fits.

\subsection{Flexible-E Model}

Recently \cite{emery_porcu_white_2022} (Theorem 3B) gave another set of sufficient conditions with the goal of alleviating the restriction imposed on $|\sigma_{ij}/(\sigma_{ii}\sigma_{jj})^{1/2}|$ by the Flexible-A model: 
\begin{enumerate}
    \item 
    $(\nu_{ij})_{i,j = 1}^p$ is conditionally negative semidefinite,
    \item For $\beta > 0$,
   $(\alpha^{-2}_{ij} - \beta\nu_{ij})_{i,j = 1}^p$ is conditionally negative semidefinite,
       \item 
$\sigma_{ij} = (\sigma_{ii}\sigma_{jj})^{1/2}V_{ij}(u_{ii}u_{jj})^{-1/2}u_{ij}$
 where
    \begin{align*}
         u_{ij} = \alpha_{ij}^{2\nu_{ij}}\beta^{\nu_{ij}}\exp(\nu_{ij})\Gamma(\nu_{ij})
 \end{align*}
 and $V$ is a correlation matrix.
\end{enumerate}
\cite{emery_porcu_white_2022} discuss in some detail how the conditions of the two models are related. A practical way to ensure that condition $1$ of the Flexible-E model holds is to parameterize $\nu_{ij}$ exactly as in condition $1$ of the Flexible-A model. We use this parameterization in our software. Similarly, we ensure $2$ holds by defining
\begin{align*}
    \alpha^{-2}_{ij}  = \frac{\alpha^{-2}_{ii} + \alpha^{-2}_{jj}}{2}  + \Delta_B(1 - B_{ij}) + \beta\Big(\nu_{ij} - \frac{  \nu_{ii} + \nu_{jj}}{2}\Big),
\end{align*}
where $\Delta_B\geq 0,$ $B$ is a positive correlation matrix, and $\beta >0$. We will refer to the model defined by these conditions as the \emph{Flexible}-$E$ model.

\subsection{Unconstrained Model}\label{sec:unconstrained}

Lastly, we also consider an unconstrained model. ``Unconstrained'' is a bit of a misnomer here because we do enforce positivity on parameters that must be positive, and we force any parameter that can be interpreted as a correlation to be between $-1$ and $1$. In particular, we use a log/exponential link for all range parameters, all smoothness parameters, marginal covariance parameters, and marginal nugget parameters. For the cross covariance and cross nugget parameters, we use
\begin{align*}
\sigma_{ij} &= \sqrt{\sigma_{ii}\sigma_{jj}} \frac{2}{\pi}\arctan(s_{ij}) \\
\tau_{ij} &= \sqrt{\tau_{ii}\tau_{jj}} \frac{2}{\pi}\arctan(t_{ij})
\end{align*}
where $s_{ij}$ and $t_{ij}$ are unconstrained. This model is unconstrained in the sense that it does not limit the flexibility of the multivariate Mat\'ern in any way. However, it may return parameters that are invalid. Aside from checking various necessary conditions, such as positivity of $\alpha_{ij}$ and $\nu_{ij}$, it is not generally possible to check necessary and sufficient conditions for validity, since these conditions have not been fully characterized.

\section{Vecchia's Approximation for Multivariate Data}\label{vecchia}

Let $\bm{y} = (y_1,\ldots,y_n)^{\top}$ be a vector of all of the responses in a multivariate spatial dataset, and let $\pi: \{1,\ldots,n\} \to \{1,\ldots,n\}$ be a permutation (reordering), so that $y_{\pi(k)}$ is the $k$th observation in a reordering of $(y_1,\ldots,y_n)$. Additionally, let $g(k)$ be a subset of $\{\ell : \ell < k\}$ and $\bm{y}_{g(k)}$ be the vector $(y_\ell : \ell \in g(k))$. For density $p$, Vecchia's approximation of the likelihood is
\begin{align*}
L(\theta; \, \bm{y}, \pi, g) = 
\prod_{k=1}^n p\Big( y_{\pi(k)} \, | \, \bm{y}_{ \pi( g(k) ) } ;\, \theta \Big).
\end{align*}
Constructing a Vecchia approximation boils down to selecting the permutation and the sequence of functions $g(1), \ldots, g(n)$, commonly referred to as the ordering and the conditioning sets. 
Usually the information about the locations is used to make these selections. For the ordering, observations can be sorted by one of the coordinates, or a space-filling ordering can be used. For the conditioning sets, typically the $m$ nearest neighbors are used.

These choices are complicated in the multivariate case, since one of the pieces of information in the data--the multivariate component--is usually categorical, and thus has no natural distance metrics upon which to base an ordering or conditioning set. We must not allow that conundrum lull us into paralysis. Our strategy here is simply to propose a few options for choosing the ordering and conditioning sets, and test them out on real datasets. 

\vspace{12pt}

\noindent \textbf{Ordering}: We consider three ordering schemes:
\begin{enumerate}
\item Completely random: $\pi$ is a random permutation of $1,\ldots,n$.
\item Order by component: observations are sorted by component, alphabetically, then ordered randomly within component.
\item Cycle through components: The components are ordered alphabetically, then the ordering repeatedly cycles through the components; in each cycle an observation is selected at random from each component.
\end{enumerate}

\vspace{12pt}

\noindent \textbf{Neighbor Selection}: We consider three neighbor-selection rules:
\begin{enumerate}
\item $m$ nearest neighbors, regardless of component
\item Ensure we select roughly $m/p$ nearest neighbors from each component.
\item If observation $k$ has component $i$, ensure we select roughly $2 m/(p+1)$ nearest neighbors from component $i$, and roughly $m/(p+1)$ nearest neighbors from the other components.
\end{enumerate}

\vspace{12pt}

We view the first option for both ordering and neighbor selection as the baseline choice because they ignore the component information. The other options make use of the component information. The two other ordering options are intended to explore whether there is any benefit to treating the components (roughly) exchangeably. The two other neighbor selection schemes are intended to explore whether there is a benefit to ensuring that every component is represented in the conditioning set, and whether, further, there is a benefit to privileging the component whose conditional distribution is being approximated.

\section{Optimization}\label{optimization}

We performed many numerical studies with different datasets and parameterizations in the process of developing our software, much of which cannot be included in the paper. This section serves the purpose of leaving some notes to assist anyone trying to improve upon our methods.

Without a reparameterization, the two-, three-, and four-component multivariate Mat\'ern models have 12, 24, and 40 parameters, respectively. Optimization with respect to these parameters is difficult due simply to the high-dimensionality of the parameter space, and more subtly, due to the fact that some parameters play similar roles in the model; for example, $\sigma_{12}$, $\alpha_{12}$, $\nu_{12}$, and $\tau_{12}$ all control dependence between the first and second components, albeit in different ways. \cite{guinness2022nonparametric} conducted a simulation study on two-, three-, and four-component models, and found that parameter estimation required thousands of Nelder-Mead iterations when using the \verb!optim! function in R. In that study, optimization with respect to an unconstrained four-component model often failed to converge, even after 6000 iterations.

Optimization performance can be improved if first and second derivative information is computed and used. \cite{guinness2021gaussian} provided formulas for the gradient and Fisher information of Vecchia's loglikelihood approximation, showing that the use of a Fisher scoring algorithm can provide vast improvements in optimization performance, especially when the model contains many parameters. 
Our Fisher scoring algorithm attempts to update the parameters with the following step:
\begin{align*}
\bm{\theta}^{k+1} = \bm{\theta}^k - \mathcal{I}(\bm{\theta}^k)^{-1} \nabla \ell( \bm{\theta}^k )
\end{align*}
where $\mathcal{I}(\bm{\theta})$ is the Fisher information matrix, and $\nabla \ell(\bm{\theta})$ is the gradient of the loglikelihood. We then check whether the loglikelihood has been improved. If not, we take a smaller step in the same direction. If the loglikelihood still has not improved we take progressively smaller steps in the direction of the gradient, and stop the optimization if we cannot increase along the gradient. These modifications to the step are important in the unconstrained case because it is easy to step into a place that produces a covariance matrix that is not positive definite. Otherwise, the optimization is stopped when the dot product between the step and the gradient is less than $10^{-4}$, which is the default stopping criterion in GpGp. To generate starting values, we use the result of an optimization over the marginal parameters. We cap the number of iterations at 40.

We found that that link function can have a large impact on the reliability of the optimization procedure. 
\cite{emery_porcu_white_2022} used an exponential/log link for the diagonals of the nugget covariance matrix, and an identity link for the off-diagonals. By fitting the model to various datasets, we found that, in the context of Fisher scoring, that link is unreliable because when the nugget is very close to zero, there is little information about the log of the marginal nugget, and a lot of information about the raw cross nuggets. In other words, we are not sure whether the log marginal nugget is $-6$ or $-10$, but we are certain that the cross nuggets are within $\exp(-6)$ of zero. This makes the Fisher information matrix poorly conditioned, and thus the optimization steps unreliable. This is a problem that one would not encounter unless working with second-derivative information. The link functions in Subsection \ref{sec:unconstrained} produced Fisher information matrices that are better-behaved. At every step, we compute the ratio between the smallest and largest eigenvalues of the Fisher information matrix (i.e.\ reciprocal of its condition number); if the ratio is too small, we set the smallest ratios to $10^{-5}$.

One of the lessons learned from the development of the GpGp optimization procedures is that Fisher scoring struggles to converge when one or more of the maximum likelihood parameters sits near--or even outside of--the boundary of the parameter space. For example, it is not difficult to simulate a univariate dataset that has a maximum likelihood nugget (in the Mat\'ern model) that is negative. When the nugget is parameterized on the log scale, the Fisher scoring algorithm tries to make the log nugget more and more negative, without ever reaching the maximum. To deal with this problem, GpGp uses minor penalties on some parameters, including the nugget, the marginal variance, and the smoothness. We use these same penalties on the marginal parameters in the multivariate Mat\'ern. We experimented with similar penalties on the other multivariate Mat\'ern parameters but did not achieve consistent success in improving the optimization by adding penalties, so we did not include them.

\section{Datasets}\label{data}
\subsection*{Weather Data}

Both \cite{gneiting2010matern} and \cite{apanasovich2012valid} analyzed a bivariate meteorological dataset. The data consist of forecast errors of pressure and temperature at 157 locations in the Pacific Northwest. \cite{gneiting2010matern} noted that pressure and temperature errors tend to be negatively correlated, which was confirmed by fitting various bivariate spatial models to the data. The data were obtained from the \verb!RandomFields! R package.

\begin{figure}     
  \centering
  \includegraphics[width=\textwidth]{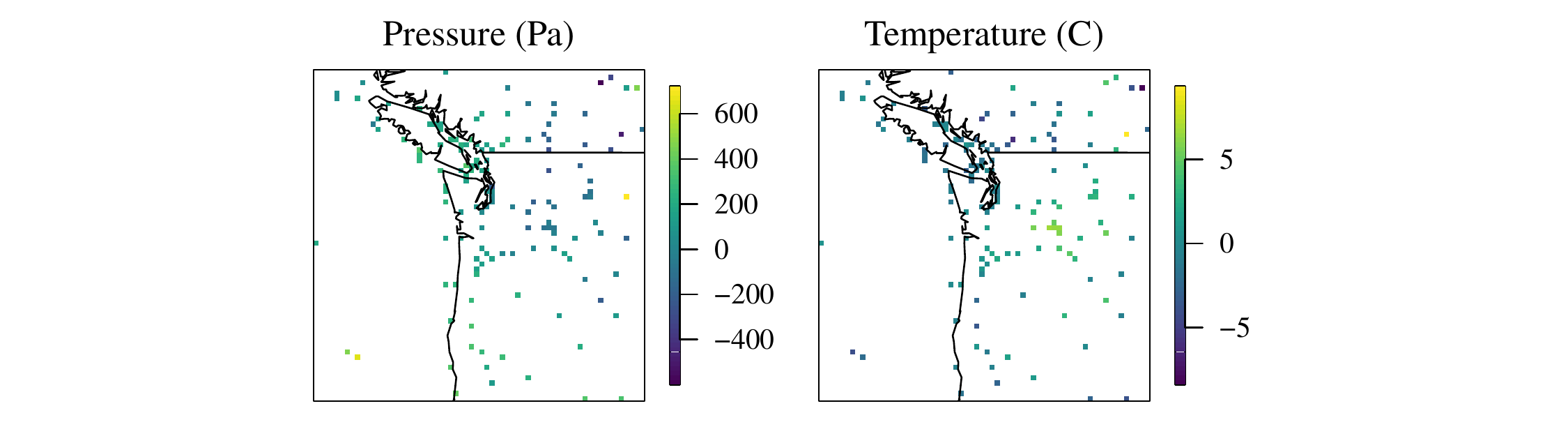}
  \caption{\label{fig:weather} Pressure and temperature data from RandomFields package at 157 locations in the Pacific Northwest.}
\end{figure}

\subsection*{French Soil Data}

\begin{figure}     
  \centering
  \includegraphics[width=\textwidth]{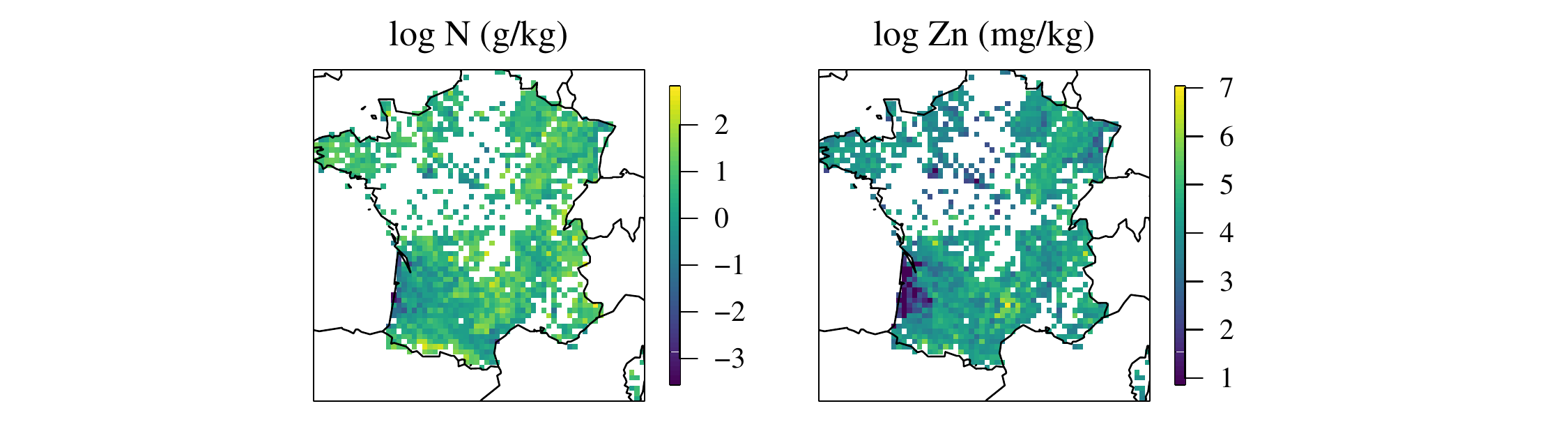}
  \caption{\label{fig:france01} Nitrogen (left) and zinc (right) in the French soil monitoring network at 1379 locations collected in 2006-2008.}
\end{figure}

The French National Soil Monitoring network is a program designed to monitor the presence of trace elements in French soils. The network consists of a 16 km $\times$ 16 km grid of locations covering French territory. Soil samples are taken as close as possible to the grid centers, and samples are measured for trace elements. We downloaded the data from \url{https://doi.org/10.15454/QSXKGA}. This dataset contains measurements of over 50 quantities with sample dates reported between June 2000 and June 2009. We subset the data to samples taken from the top soil layer in 2006, 2007, and 2008, which includes 1379 individual sample locations. A multivariate spatial principal components analysis of 8 of the elements is reported in \cite{saby2009multivariate}. We use components \verb!n_tot_31_1! and \verb!zn_tot_hf!, which are measurements of nitrogen and zinc. We omitted duplicated locations. These components are plotted in Figure \ref{fig:france01}.

\subsection*{KAUST Competition Data}

\begin{figure}
\centering
\includegraphics[width=0.6\textwidth]{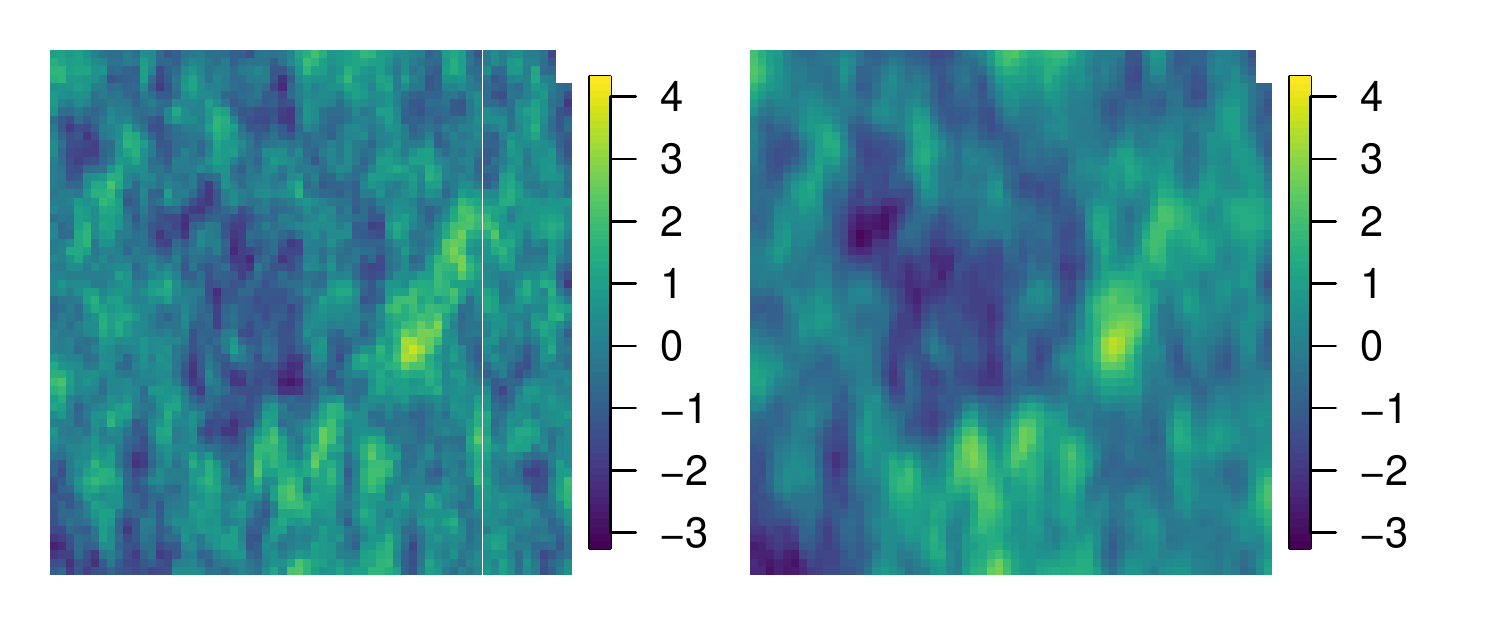}
\caption{\label{fig:competition} Bivariate dataset used in the 2022 KAUST competition}
\end{figure}

The Spatio-Temporal Statistics and Data Science group at KAUST has run two spatial data analysis competitions in which teams are scored on prediction metrics on various spatial datasets. The second competition, run in 2022, included spatial-temporal and bivariate spatial multivariate datasets. Here, we analyze the bivariate spatial data from Competition 3a, specifically the file \verb!3a_1_train.csv! which can be obtained from \url{https://www.kaggle.com/competitions/2022-kaust-ss-competition-3a/data}. A preliminary version of the methods in this paper were used in the competition, and that version finished in first place in the scoring metrics, though several teams finished in close proximity. The data are plotted in Figure \ref{fig:competition}.

\subsection*{Bangladesh Water Quality Data}

\begin{figure}     
  \centering
  \includegraphics[width=\textwidth]{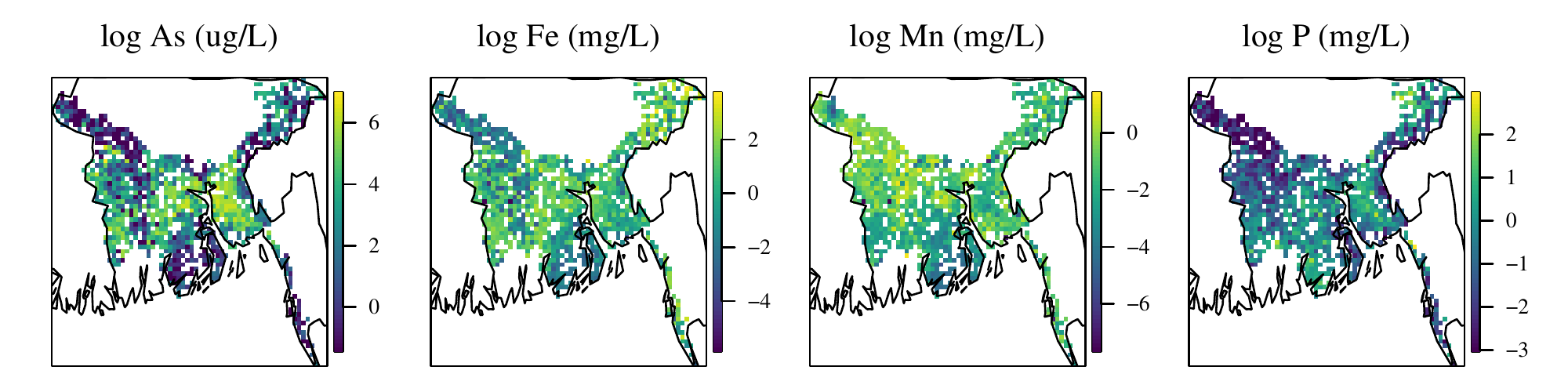}
  \caption{\label{fig:bang} Bangladesh Water quality data samples from 1998, including arsenic, iron, manganese, and phosphorous}
\end{figure}

In response to the detection of arsenic in Western Bangladesh, the British Geological Survey (BGS) conducted a large-scale well-water sampling program throughout Bangladesh in 1998 and 1999 \citep{kinniburgh2001arsenic}. The BGS sampled water from over 3500 wells, and tested them for arsenic and various other elements. All of the data can be downloaded from \url{https://www2.bgs.ac.uk/groundwater/health/arsenic/Bangladesh/data.html}; we specifically analyze the arsenic, iron, manganese, and phosphorous data collected in 1998, from the DPHE/BGS National Hydrochemical Survey. We omitted duplicated locations. The data are plotted in Figure \ref{fig:bang}.
\subsection*{GOES 16 Radiance Data}

\begin{figure}
\centering
\includegraphics[width=\textwidth]{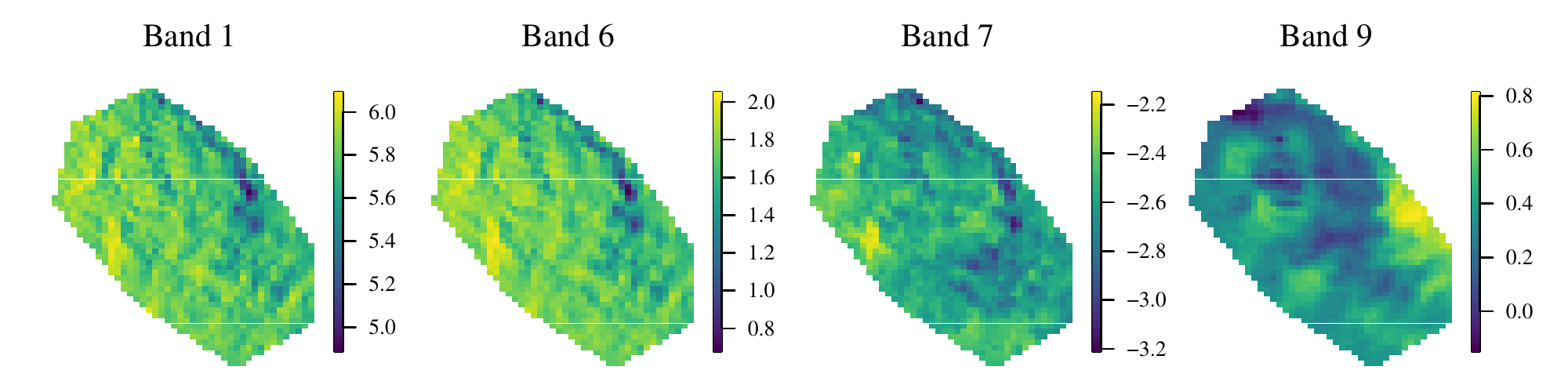}
\caption{\label{fig:goes} Log radiances from bands 1, 6, 7, 9 from the Hurricane Florence data collected by the GOES-16 Advanced Baseline Imager.}
\end{figure}

\cite{guinness2022nonparametric} analyzed data from the advanced baseline imager on the GOES-16 satellite, which operates in geostationary orbit. The imager records reflected radiances in 16 wavelength bands at a spatial resolution of up to 1km and a temporal resolution up to 1 minute. We analyze the Hurricane Florence data introduced in \cite{guinness2022nonparametric}, which includes bands 1, 6, 7, and 9. Specifically, we analyze data from the 30th minute of the Florence data used in that paper. Data were extracted from an outer edge of the hurricane cloud, selected visually for stationarity. The data are plotted in Figure \ref{fig:goes}.

\section{Numerical Results}\label{results}

All computations are done on an 8-core (16-thread) Intel(R) Xeon(R) W-2145 CPU @ 3.70GHz. Reported loglikelihoods are Vecchia approximations, and aside from the first study, all use the completely random ordering scheme, and nearest neighbors that ignore the component. We did not have room in most of the tables to include timing. For the French soil, the Bangladesh water, and the GOES-16 data, one iteration takes roughly 1 second, so nearly all of them complete within 1 minute. The weather dataset is smaller, so runs much faster. The fits to the competition data--which has 90{,}000 observations--take roughly 5 minutes each.

\subsection{Effect of Vecchia's Approximation}

Our first numerical study explores the effect of ordering, conditioning, and the number of neighbors in Vecchia's approximation. We fit the independent, parsimonious, Flexible-A, Flexible-E, and unconstrained models to the French Soil Data under all combinations of the proposed orderings, conditioning rules, and 20 and 40 neighbors. The results are in Table \ref{tab:vecchia}. In short, our results are somewhat inconclusive in terms of selecting a clear winner, though it seems that there is not much to be lost by ignoring the components when computing the ordering and conditioning sets (setting rand-any). Our experience is that Vecchia's approximation is reasonably accurate on spatial datasets, so we interpret these results to mean that the ordering and conditioning choices do not have a large impact on this bivariate spatial dataset. Certainly more study is needed to identify scenarios where the ordering and conditioning has a larger impact, especially with different datasets and more components. For the rest of the numerical results, we use the rand-any setting with 20 neighbors. We view the rand-any setting as a ``hands-off'' setting that is unlikely to introduce a systematic issue.

In terms of model comparison, the dependent models give significantly higher likelihoods than the independent model, and there are relatively small differences among the dependent models. This is consistent with findings in the literature on other datasets. In numerical studies to follow, we further explore differences between the various models.

\begin{table}[t]
\caption{\label{tab:vecchia} Results of effect of choices in Vecchia's approximation for the various models fit to the French soil data. The reported loglikelihoods are differences from the maximum in the table. Bolded entry is the highest within each combination of model and $m$ setting. rand = completely random ordering, comp = order by component, cyc = order cycling through components, any = nearest neighbors regardless of component, bal = attempt to select an equal number of neighbors from each component, pref = select more neighbors from the current component.}
\centering
{\footnotesize
\begin{tabular}{lrrrrrrrrrrr}
 && \multicolumn{2}{c}{Independent} & \multicolumn{2}{c}{Parsimonious} & \multicolumn{2}{c}{Flexible-A} & \multicolumn{2}{c}{Flexible-E} & \multicolumn{2}{c}{Unconstrained} \\
  \hline
Setting & $m$ & loglik & iter & loglik & iter & loglik & iter & loglik & iter & loglik & iter \\ 
  \hline
  rand-any & 20 & $-$218.00 & 27 & $-$14.20 & 26 & $-$11.13 & 40 & $-$8.79 & 40 & \textbf{$-$3.56} & 40 \\ 
  comp-any & 20 & $-$243.12 & 12 & $-$19.02 & 31 & $-$15.87 & 40 & $-$14.99 & 40 & $-$14.61 & 40 \\ 
  cyc-any & 20 & $-$229.09 & 19 & $-$18.75 & 24 & $-$12.88 & 40 & $-$12.76 & 40 & $-$11.28 & 40 \\ 
  rand-bal & 20 & $-$218.95 & 24 & $-$13.35 & 27 & \textbf{$-$9.29} & 40 & $-$7.49 & 40 & $-$3.62 & 40 \\ 
  comp-bal & 20 & $-$222.49 & 8 & $-$11.38 & 8 & $-$17.81 & 40 & $-$8.00 & 40 & $-$7.13 & 40 \\ 
  cyc-bal & 20 & $-$227.41 & 38 & $-$18.55 & 27 & $-$13.09 & 40 & $-$13.10 & 40 & $-$11.45 & 12 \\ 
  rand-pref & 20 & \textbf{$-$214.42} & 22 & $-$11.48 & 24 & $-$11.67 & 40 & \textbf{$-$7.43} & 40 & $-$5.60 & 25 \\ 
  comp-pref & 20 & $-$219.38 & 8 & \textbf{$-$10.55} & 7 & $-$29.23 & 40 & $-$9.42 & 40 & $-$8.64 & 40 \\ 
  cyc-pref & 20 & $-$221.64 & 23 & $-$19.23 & 13 & $-$15.26 & 40 & $-$13.68 & 40 & $-$11.66 & 37 \\ 
  \hline
  rand-any & 40 & $-$214.80 & 17 & $-$12.11 & 16 & $-$9.74 & 40 & $-$5.71 & 40 & $-$3.84 & 40 \\ 
  comp-any & 40 & $-$221.67 & 12 & $-$15.43 & 16 & $-$10.14 & 40 & $-$8.30 & 40 & $-$6.63 & 40 \\ 
  cyc-any & 40 & $-$214.39 & 8 & $-$11.76 & 23 & $-$9.03 & 40 & $-$5.79 & 40 & $-$4.68 & 11 \\ 
  rand-bal & 40 & \textbf{$-$213.79} & 12 & $-$11.60 & 15 & $-$12.01 & 40 & $-$5.61 & 40 & $-$3.47 & 40 \\ 
  comp-bal & 40 & $-$215.50 & 8 & $-$11.26 & 8 & $-$5.62 & 40 & $-$4.99 & 40 & $-$3.05 & 40 \\ 
  cyc-bal & 40 & $-$215.17 & 8 & $-$12.62 & 21 & $-$7.39 & 40 & $-$6.31 & 40 & $-$3.56 & 12 \\ 
  rand-pref & 40 & $-$213.92 & 8 & $-$13.26 & 16 & $-$6.60 & 40 & $-$6.81 & 40 & $-$1.87 & 9 \\ 
  comp-pref & 40 & $-$214.80 & 7 & \textbf{$-$5.52} & 8 & \textbf{$-$2.10} & 40 & \textbf{$-$1.84} & 36 & \textbf{0.00} & 40 \\ 
  cyc-pref & 40 & $-$214.62 & 8 & $-$12.38 & 8 & $-$5.70 & 40 & $-$5.85 & 40 & $-$2.15 & 10 \\ 
  \hline
\end{tabular}
}
\end{table}

\subsection{Comparison of Models}

We fit the bivariate Mat\'ern model to the weather data, the French soil data, and the competition data under the independent, parsimonious, Flexible-A, Flexible-E, and unconstrained models. The estimated parameters and loglikelihoods are in Table \ref{tab:model}. There is a roughly 10 point loglikelihood increase between the independent and various dependent models fit to the weather data, which has 157 locations.
The dependent models provide significantly better fits to the French soil data, which is a larger dataset of 1379 locations. There are some minor differences in loglikelihoods among the dependent models; the ordering from best to worst is unconstrained, Flexible-E, Flexible-A, then parsimonious. An inspection of the parameters for the Flexible-E model versus the unconstrained model reveals that the sizes of the marginal nuggets change. Interestingly, the smoothness parameters are small and change as well. Similarly, in the weather data, though our loglikelihoods are close to those reported in \cite{apanasovich2012valid}, the parameter estimates are somewhat different. We believe this is due to the inclusion of the cross nugget, since we get similar estimates to those reported in \cite{apanasovich2012valid} when we set the cross nugget to zero. The relationship between nuggets and the other parameters is explored further in the next subsection.

\begin{table}[ht]
\caption{\label{tab:model} Parameter estimates and loglikelihoods for the weather, French soil, and competition datasets.}
\centering
{\footnotesize
\begin{tabular}{llrrrrr}
Dataset & & Independent & Parsimonious & Flexible-A & Flexible-E & Unconstrained \\ 
\hline
Weather & $\sigma_{11}$ & 53393.48 & 47677.52 & 47696.49 & 47686.68 & 49394.45 \\ 
   & $\sigma_{12}$ & 0.00 & $-$289.64 & $-$268.12 & $-$268.08 & $-$291.84 \\ 
   & $\sigma_{22}$ & 6.76 & 6.91 & 6.70 & 6.70 & 6.70 \\ 
   & $\alpha_{11}$ & 59.05 & 93.66 & 145.96 & 146.63 & 127.04 \\ 
   & $\alpha_{12}$ & 65.38 & 93.66 & 95.02 & 94.36 & 130.08 \\ 
   & $\alpha_{22}$ & 94.90 & 93.66 & 75.69 & 74.94 & 71.24 \\ 
   & $\nu_{11}$ & 2.52 & 1.18 & 0.65 & 0.64 & 0.83 \\ 
   & $\nu_{12}$ & 1.93 & 0.89 & 0.68 & 0.68 & 0.55 \\ 
   & $\nu_{22}$ & 0.58 & 0.60 & 0.71 & 0.72 & 0.75 \\ 
   & $\tau_{11}$ & 4807.41 & 4108.02 & 2399.04 & 2313.08 & 3262.89 \\ 
   & $\tau_{12}$ & 0.00 & 6.34 & 12.78 & 12.97 & 17.02 \\ 
   & $\tau_{22}$ & 0.00 & 0.01 & 0.07 & 0.07 & 0.09 \\ 
   & loglik & $-$1273.50 & $-$1264.33 & $-$1263.62 & $-$1263.61 & $-$1263.19 \\ 
   \hline
  French Soil & $\sigma_{11}$ & 0.59 & 0.45 & 0.48 & 0.44 & 0.37 \\ 
   & $\sigma_{12}$ & 0.00 & 0.31 & 0.31 & 0.29 & 0.40 \\ 
   & $\sigma_{22}$ & 0.64 & 0.43 & 0.54 & 0.50 & 0.79 \\ 
   & $\alpha_{11}$ & 3.50 & 1.10 & 1.49 & 1.23 & 0.80 \\ 
   & $\alpha_{12}$ & 1.55 & 1.10 & 1.66 & 1.22 & 2.47 \\ 
   & $\alpha_{22}$ & 3.33 & 1.10 & 1.92 & 1.59 & 7.72 \\ 
   & $\nu_{11}$ & 0.14 & 0.15 & 0.16 & 0.17 & 0.46 \\ 
   & $\nu_{12}$ & 0.51 & 0.22 & 0.29 & 0.38 & 0.34 \\ 
   & $\nu_{22}$ & 0.14 & 0.29 & 0.22 & 0.25 & 0.14 \\ 
   & $\tau_{11}$ & 0.01 & 0.01 & 0.02 & 0.04 & 0.17 \\ 
   & $\tau_{12}$ & 0.00 & 0.00 & 0.05 & 0.07 & 0.07 \\ 
   & $\tau_{22}$ & 0.02 & 0.16 & 0.13 & 0.14 & 0.03 \\ 
   & loglik & $-$2370.62 & $-$2166.82 & $-$2163.75 & $-$2161.41 & $-$2156.24 \\ 
   \hline
  Competition & $\sigma_{11}$ & 0.96 & 0.95 & 0.96 & 0.96 & 0.96 \\ 
   & $\sigma_{12}$ & 0.00 & 0.82 & 0.84 & 0.84 & 0.84 \\ 
   & $\sigma_{22}$ & 0.95 & 1.03 & 1.07 & 1.06 & 1.06 \\ 
   & $\alpha_{11}$ & 0.03 & 0.03 & 0.03 & 0.03 & 0.03 \\ 
   & $\alpha_{12}$ & 0.03 & 0.03 & 0.03 & 0.03 & 0.03 \\ 
   & $\alpha_{22}$ & 0.03 & 0.03 & 0.03 & 0.03 & 0.03 \\ 
   & $\nu_{11}$ & 0.60 & 0.60 & 0.61 & 0.61 & 0.61 \\ 
   & $\nu_{12}$ & 1.38 & 1.00 & 1.01 & 1.01 & 1.01 \\ 
   & $\nu_{22}$ & 1.41 & 1.40 & 1.40 & 1.40 & 1.41 \\ 
   & $\tau_{11}$ & 0.00 & 0.00 & 0.00 & 0.00 & 0.00 \\ 
   & $\tau_{12}$ & 0.00 & 0.00 & 0.00 & 0.00 & 0.00 \\ 
   & $\tau_{22}$ & 0.00 & 0.00 & 0.00 & 0.00 & 0.00 \\ 
   & loglik & 90809.25 & 118953.00 & 118956.37 & 118956.22 & 118956.59 \\ 
  \hline
\end{tabular}
}
\end{table}

\subsection{Effect of Correlated Nuggets}

One of our most interesting findings is that forcing zero dependence across components in the nugget can seriously distort the estimates of the range and smoothness parameters. This is demonstrated in Table \ref{tab:corrnug}, where we show parameter estimates for various combinations
of elemental components in the Bangladesh data, under the Flexible-E model. When the cross nuggets are forced to be zero, the estimated smoothness parameters can become close to zero. This has an appreciable effect on the maximum likelihood, on the order of 10-50 points, and usually has an impact on the smoothness and range parameters.

\begin{figure}
\centering
\includegraphics[width=\textwidth]{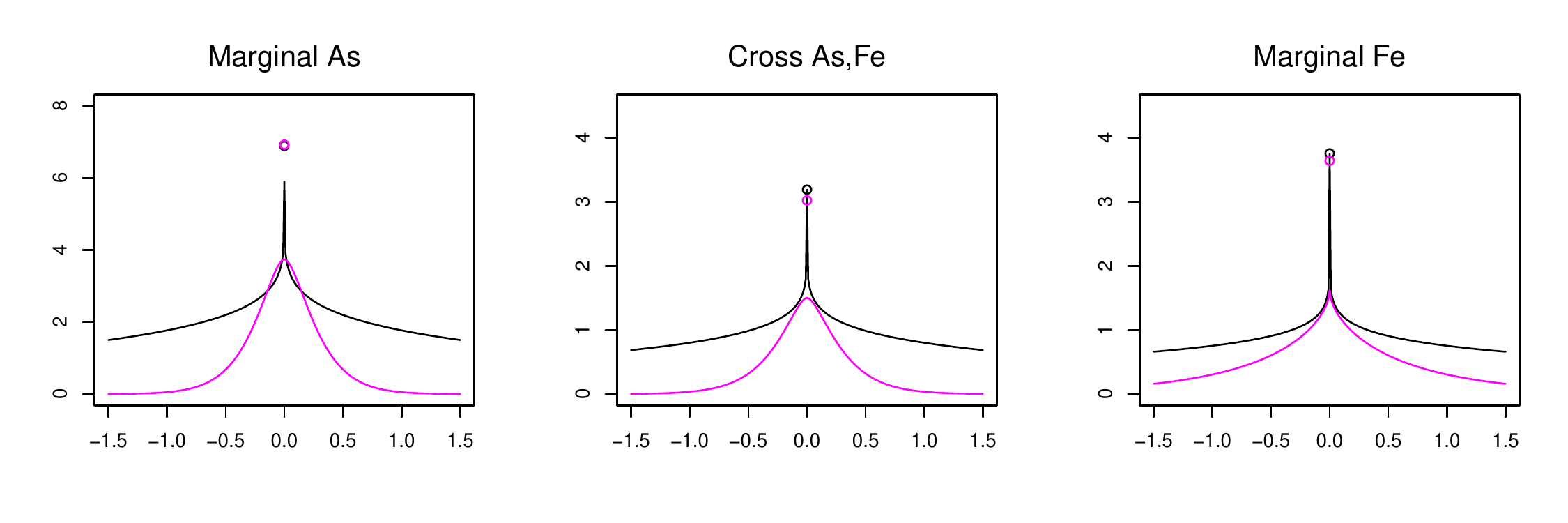}
\caption{\label{fig:corrnug} Estimated marginal and cross covariance functions  for the As and Fe components in the Bangladesh dataset with the restriction that $\tau_{12} = 0$ (black) and without (magenta).}
\end{figure}

Figure \ref{fig:corrnug} shows a plausible explanation for the behavior seen in Table \ref{tab:corrnug}. The middle panel shows the estimated cross covariance functions with and without the restriction that the cross nugget must be zero. When the cross nugget is restricted to be zero (black), the model can squeeze itself into pretending it has a correlated nugget by making the smoothness small, creating a spike at zero. This comes at the cost of limiting the shape of the Mat\'ern function. By contrast, when the cross nugget is allowed to be non-zero (magenta), the Mat\'ern function retains its flexibility.

\begin{table}
  \caption{\label{tab:corrnug} Results for allowing correlated nuggets, vs forcing zero correlation, on the Bangladesh water quality data, using the Flexible-E model. For each pair of components, the first row forces $\tau_{12} = 0$, whereas the second row allows it to be non-zero. As = arsenic, Fe = iron, Mn = manganese, P = phosphorous.}
\centering
{\small
\begin{tabular}{lrrrrrrrrrrrr}
Comp & $\sigma_{11}$ & $\sigma_{12}$ & $\sigma_{22}$ & $\alpha_{11}$ & $\alpha_{12}$ & $\alpha_{22}$ & $\nu_{11}$ & $\nu_{12}$ & $\nu_{22}$ & $\tau_{11}$ & $\tau_{12}$ & $\tau_{22}$ \\ 
  \hline
  As,Fe & 5.90 & 3.19 & 3.75 & 8.20 & 10.25 & 16.93 & 0.08 & 0.06 & 0.04 & 0.99 & 0.00 & 0.01 \\ 
   & 3.73 & 1.50 & 1.62 & 0.16 & 0.20 & 0.88 & 1.42 & 1.12 & 0.28 & 3.19 & 1.52 & 2.03 \\ 
   \hline
  As,Mn & 3.75 & $-$0.04 & 1.10 & 0.15 & 0.17 & 0.21 & 1.65 & 1.59 & 0.79 & 3.22 & 0.00 & 1.23 \\ 
   & 3.89 & $-$0.25 & 1.14 & 0.18 & 0.20 & 0.23 & 1.29 & 1.00 & 0.70 & 3.18 & 0.28 & 1.21 \\ 
   \hline
  As,P & 6.23 & 2.28 & 1.90 & 13.24 & 15.67 & 20.45 & 0.07 & 0.07 & 0.06 & 0.88 & 0.00 & 0.01 \\ 
   & 4.16 & 1.54 & 1.24 & 0.16 & 0.17 & 0.47 & 1.66 & 1.51 & 0.49 & 3.23 & 0.92 & 0.77 \\ 
   \hline
  Fe,Mn & 3.98 & 0.61 & 2.51 & 122.70 & 0.00 & 209.57 & 0.03 & 0.03 & 0.04 & 0.03 & 0.00 & 0.07 \\ 
   & 1.64 & 0.12 & 1.10 & 0.49 & 0.13 & 0.29 & 0.45 & 7.72 & 0.64 & 2.14 & 0.59 & 1.23 \\ 
   \hline
  Fe,P & 3.86 & 1.43 & 1.70 & 45.82 & 42.06 & 39.16 & 0.03 & 0.05 & 0.06 & 0.01 & 0.00 & 0.22 \\ 
   & 1.59 & 0.70 & 1.09 & 0.50 & 0.34 & 0.28 & 0.49 & 0.95 & 0.82 & 2.19 & 0.68 & 0.82 \\ 
   \hline
  Mn,P & 0.99 & $-$0.31 & 0.97 & 0.17 & 0.12 & 0.25 & 1.03 & 4.38 & 0.82 & 1.26 & 0.00 & 0.82 \\ 
   & 1.00 & $-$0.34 & 0.99 & 0.21 & 0.23 & 0.27 & 0.79 & 0.77 & 0.76 & 1.24 & 0.15 & 0.81 \\ 
   \hline
\end{tabular}
}
\end{table}

\subsection{Many Datasets}

Lastly, we fit the various multivariate Mat\'ern models to all combinations of components in the Bangladesh and GOES data. The results are in Table \ref{tab:many_datasets}. The reported times for each fit are in minutes, and the reported loglikelihoods are differences between each model and the independent model on the same dataset. For the Bangladesh Water data, manganese (Mn) appears to be the least correlated with the other components, in that the improvement over the independent model is small when a single element is paired with Mn. There is generally little difference between the different dependent models on the Bangladesh Water data. On the GOES data, there is strong dependence between components 1 and 6; whenever they are paired together, the likelihood is more than 1000 units higher than the independent model. This is consistent with a visual inspection of Figure \ref{fig:goes}. There are a few cases showing differences of more than 10 loglikelihood units between the different dependent models, particularly in the four-component model. Many of the fits ran into the upper limit of 40 iterations, which could explain the differences, but also highlights the difficulty of optimizing the likelihood over high-dimensional parameter spaces.

\begin{table}
\caption{\label{tab:many_datasets} Results of fitting the various model parameterizations to all combinations of the Bangladesh water data (first 11 rows, included components identified by elemental abbreviations, As, Fe, Mn, and P) and the GOES-16 radiance data (last 11 rows, included components identified by GOES-16 band numbers, 1, 6, 7, and 9). Within each row, the loglikelihoods are the differences above the independent model fit to the same dataset. In every case, the Unconstrained fit provides the highest loglikelihood. Times are in minutes.}
\centering
{\footnotesize
\begin{tabular}{lrrrrrrrrrrrr}
& \multicolumn{3}{c}{Parsimonious}&\multicolumn{3}{c}{Flexible-A} &\multicolumn{3}{c}{Flexible-E} &\multicolumn{3}{c}{Unconstrained}  \\
  \hline
Comp & loglik & iter & time & loglik & iter & time & loglik & iter & time & loglik & iter & time \\ 
  \hline
  As,Fe & 387.84 & 35 & 0.4 & 388.29 & 40 & 0.9 & 392.14 & 40 & 0.6 & 392.22 & 16 & 0.3 \\ 
  As,Mn & 15.84 & 16 & 0.2 & 15.85 & 24 & 0.3 & 15.85 & 24 & 0.3 & 18.21 & 40 & 1.3 \\ 
  As,P & 364.70 & 40 & 0.5 & 365.90 & 40 & 0.8 & 369.89 & 40 & 0.5 & 370.09 & 40 & 0.5 \\ 
  Fe,Mn & 106.83 & 11 & 0.1 & 108.35 & 22 & 0.3 & 108.48 & 20 & 0.4 & 108.50 & 40 & 1.1 \\ 
  Fe,P & 262.57 & 13 & 0.2 & 263.20 & 16 & 0.2 & 263.19 & 17 & 0.3 & 263.45 & 33 & 0.4 \\ 
  Mn,P & 14.70 & 17 & 0.2 & 14.72 & 31 & 0.5 & 14.72 & 31 & 0.5 & 17.19 & 15 & 0.2 \\ 
  As,Fe,Mn & 493.02 & 26 & 0.5 & 496.08 & 27 & 0.5 & 496.16 & 27 & 0.6 & 498.69 & 40 & 2.0 \\ 
  As,Fe,P & 805.69 & 10 & 0.3 & 811.74 & 40 & 0.8 & 812.00 & 40 & 1.0 & 812.63 & 8 & 0.2 \\ 
  As,Mn,P & 393.24 & 34 & 0.7 & 398.38 & 40 & 0.9 & 398.76 & 40 & 1.2 & 400.89 & 40 & 2.0 \\ 
  Fe,Mn,P & 393.17 & 22 & 0.4 & 395.78 & 40 & 1.0 & 395.95 & 40 & 0.8 & 398.75 & 21 & 0.9 \\ 
  As,Fe,Mn,P & 941.96 & 21 & 0.6 & 948.09 & 37 & 1.0 & 948.35 & 40 & 1.4 & 951.60 & 40 & 2.7 \\ 
  \hline
  1,6 & 1706.89 & 40 & 0.5 & 1722.90 & 40 & 0.5 & 1722.91 & 40 & 0.5 & 1723.10 & 40 & 0.5 \\ 
  1,7 & 534.26 & 15 & 0.2 & 531.96 & 40 & 0.9 & 539.45 & 15 & 0.2 & 539.45 & 1 & 0.0 \\ 
  1,9 & 6.92 & 11 & 0.1 & 8.65 & 40 & 0.5 & 8.60 & 40 & 0.5 & 8.70 & 12 & 0.2 \\ 
  6,7 & 653.04 & 17 & 0.2 & 656.60 & 40 & 0.9 & 659.23 & 22 & 0.4 & 659.23 & 1 & 0.0 \\ 
  6,9 & 15.32 & 14 & 0.2 & 28.81 & 40 & 0.7 & 28.82 & 36 & 0.4 & 28.89 & 10 & 0.1 \\ 
  7,9 & $-$2.98 & 14 & 0.2 & 5.59 & 40 & 0.5 & 5.66 & 40 & 0.5 & 5.80 & 27 & 0.2 \\ 
  1,6,7 & 2358.91 & 40 & 0.7 & 2376.02 & 40 & 0.9 & 2379.74 & 40 & 0.9 & 2390.47 & 40 & 0.7 \\ 
  1,6,9 & 1758.79 & 40 & 0.8 & 1803.79 & 40 & 0.7 & 1803.80 & 40 & 0.7 & 1806.87 & 40 & 0.6 \\ 
  1,7,9 & 537.47 & 35 & 0.7 & 546.70 & 40 & 0.9 & 546.68 & 40 & 0.9 & 549.27 & 40 & 0.8 \\ 
  6,7,9 & 675.60 & 40 & 0.7 & 693.74 & 40 & 1.7 & 695.40 & 40 & 0.9 & 697.58 & 29 & 0.5 \\ 
  1,6,7,9 & 2462.00 & 40 & 0.9 & 2501.30 & 40 & 1.1 & 2504.86 & 40 & 1.2 & 2516.99 & 40 & 0.9 \\ 
   \hline
\end{tabular}
}
\end{table}

\section{Discussion}

Our main contribution is the implementation of the multivariate Mat\'ern and some of the proposed parameterizations in software that allows for fast fitting to large datasets via Vecchia's approximation and optimization with Fisher scoring. A major success of our work is the ability to fit unconstrained 40-parameter models to the Bangladesh Water data and the GOES data. The software allows for the exploration of the practical implications of using the various parameterizations on real datasets. Pending feedback from this article, these methods will be pushed into the GpGp R package and made publicly available on the Comprehensive R Archive Network (CRAN).

Here are the main takeaways from our studies. Consistent with other authors, we found that there is usually little difference between the Flexible-A, Flexible-E, and unconstrained fits. Of course, we have only tried the models on the datasets presented here, and there may be some datasets for which there is a larger difference. We hope that our software will provide the means for such explorations by us and other researchers. Perhaps our most interesting finding is the effect of forcing the nugget terms to be uncorrelated across components. If the data has small-scale cross-component correlation, the multivariate Mat\'ern is capable of squeezing such correlation into the model by making the smoothness parameters small, creating a spike at zero distance. We suggest allowing the nuggets to be correlated, which frees the smoothness parameters to do their job of modeling the smoothness of the spatially correlated field.

More work is needed to fine-tune the various settings that control the Fisher scoring algorithm. Some of the issues include: how much to limit the size of the steps, how to modify the Fisher information matrix when it is poorly-conditioned, and how to modify the step when the step decreases the loglikelihood. In addition, there are probably some advantages to be gained by improving the link functions to make the Fisher information matrices better behaved. Another fruitful avenue is the development of penalties (or similarly, Bayesian priors) that push the parameters away from treacherous parts of the parameter space without severely limiting the flexibility of the model. It would also be interesting to explore the implementation of asymmetric multivariate models, including the model in \cite{li2011approach} and the more flexible model in \cite{qadir2021flexible}. Lastly, extensions to other multivariate models besides the Mat\'ern would enable model comparison studies.

\section*{Supplementary Material}

The datasets and code used for this project can be found at \url{https://github.com/yf297/GpGp_multi_paper}.






\section*{Acknowledgement}
Nicholas Saby for link to French Soil Data. KAUST Spatio-Temporal Statistics and Data Science group for the competition data. 

\section*{Funding}
This work used the Extreme Science and Engineering Discovery Environment (XSEDE), which is supported by National Science Foundation grant number ACI-1548562. This work is supported by the National Science Foundation Division of Mathematical Sciences under grant numbers 1916208 and 1953088.

\bibliographystyle{agsm}
\bibliography{refs}
\end{document}